\documentclass[a4,aps,showpacs,amsmath,floatfix]{revtex4}
\usepackage{graphicx}
\usepackage{color}
\usepackage{amsmath}
\usepackage{here}
\usepackage[utf8]{inputenc}
\usepackage{float}

\begin{document}
	\title{Spatiotemporal vortex rings in atomic Bose-Einstein condensates}
	\author{O.G. Chelpanova, Y.I. Kuriatnikov, S. Vilchinskii,  A.I. Yakimenko}
	
	\affiliation{ Department of Physics, Taras Shevchenko National University of Kyiv, 64/13, Volodymyrska Street, Kyiv 01601, Ukraine}
	
	\newenvironment{comment}{}{}
	\begin{abstract}
		 We investigate spatiotemporal vortex rings with phase dislocation both in space and time.  It is demonstrated that these structures naturally appear as a periodical in time edge phase dislocation at the low-density region of a perturbed atomic Bose-Einstein condensate. The condition of formation, dynamics and stability of the spatiotemporal vortex rings are investigated for repulsive and attractive interatomic interactions.   These theoretical findings open up a perspective for experimental observation of novel type of topological coherent structures in ultracold gases.\\[0.5cm]
		{\bf Key words:} atomic Bose-Einstein condensates,  vortices, solitons
	\end{abstract}
	
	\pacs{03.75.Kk, 05.45.Yv, 05.30.Jp} \maketitle
	
	\section{Introduction}
	
	

Formation of vortex rings is a universal phenomenon observed in classical and quantum fluids of different nature. 
In quantum liquids the well known example is  quantized vortex rings in superfluid helium, discovery of which  has greatly increased interest in vortex rings with very thin cores. 
Vortex rings play a crucial role in the decay of superflow and in quantum turbulence in helium liquids and atomic Bose-Einstein condensates (BECs)
 \cite{Tsubota13}. 
While rapid progress has been made in the theoretical studies  \cite{Ginsberg2005, Shomroni2009, Jackson1999, Pinsker2013, Ruostekoski2001} of quantized vortex rings, there has not been comparable progress in laboratory studies of vortex rings. Difficulties ensue not only with experimental detection of the vortex rings in the condensate bulk but also with their unstable evolution in trapped atomic  BECs. As the result of this instability vortex rings in realistic inhomogeneous BECs either drift to an edge of the condensate, where they decay into elementary excitations, or shrink and annihilate within the condensate bulk. 
In  work \cite{PRA13} we have suggested an experimentally feasible trapping configuration (optical tweezers) that can be used for creation,  stabilization, and manipulation of a vortex ring in a controllable and nondestructive manner. Using rotating trap of similar geometry it is possible to stabilize even more complex topological structures known as Hopf solitons \cite{Hopfion}. 

Further  prospective ways for generation novel type of vortex rings in atomic BECs can be inspired by recent findings in physics of optical vortices. 
A remarkable example of spontaneous vortex ring nucleation has been  revealed in self-saturating \cite{Desyatnikov2012} and nonlocal \cite{Arxiv17} optical nonlinear media. Formation of the vortex rings at the periphery of the wave beam in optical media is a consequence	of the nonlinear phase accumulation between the soliton peak and its tail. As was
 discovered in Ref. \cite{Desyatnikov2012} phase singularities nucleate	if this phase difference reaches the value of $\pi$ during propagation along  $z$ axis. Optical vortex rings (in  contrast to vortex rings in BECs) are static in time and appear when nonlinear phase of the self-trapped light beam breaks the wave front into a sequence of optical vortex loops around the perturbed fundamental soliton.
The spatial optical vortex is associated with region of space with field null line around which electromagnetic energy density circulates. But in general, beam propagation is not described only by its spatial features, and  temporal evolution of the wave beam parameters should be taken into account as well. As known, dynamics of optical pulsed beams with spatiotemporal amplitude and phase modulation is governed by diffraction and dispersion effects. This provides a new insight on the spatiotemporally localized waves, called \emph{
spatiotemporal vortices}, drawing on analogy with spatial screw phase dislocations \cite{Sukhorukov2005}. Spatiotemporal optical vortices have been theoretically predicted \cite{Sukhorukov2005, Bliokh2012} and very recently experimentally observed \cite{Jhajj2016} during self-focusing collapse of wave beam in air.

 One of the goal of this  work is to find the conditions for formation of spatiotemporal vortex rings and describe their general properties in atomic BEC.  Our idea is based on the well known analogy \cite{KivsharBook} between two mathematical models: nonlinear Schr\"{o}dinger equation describing propagation in the $z$ direction of the wave beam in nonlinear optical media and Gross-Pitaevskii equation (GPE) describing evolution in time $t$ of BEC wave function. It would appear reasonable that a sequence of \textit{spatiotemporal} vortex loops arises around the perturbed matter-wave soliton similar to a sequence of spatial vortex rings around perturbed optical soliton \cite{Desyatnikov2012}.  These matter wave spatiotemporal vortex rings should appear as a periodically forming circular edge phase dislocation at the periphery of the condensate cloud.

	
	The paper is organized as follows. In Sec. \ref{Sec_Model}, we describe
 model. We take into account both two-particle and three-particle interactions which can be described by GPE  with cubic-quintic nonlinearity.  In Sec. \ref{Sec_Trapped} we study dynamics of radially perturbed two-dimensional BEC in  a harmonic trap and investigate a possibility of spatiotemporal vortex ring nucleation. It turns out that these structures are not observable in trappped BEC due to sharp decay of the condensate density at the periphery of condensate. In Sec. \ref{Sec_qubic_quintic} we demonstrate that  spatiotemporal vortex rings can be observed in a condensate with attractive two-particle and repulsive three-particle interactions without external potential.

	\section{Model}\label{Sec_Model}
Dynamical properties of ultracold dilute atomic BECs can be accurately described by the mean-field Gross-Pitaevskii equation \cite{Pitaevskii2016}: 
	\begin{equation}
	\label{eq:zero}
	i\hbar \dfrac{\partial \Psi}{\partial t}=\left(-\dfrac{\hbar^2}{2 M}\nabla^2 +V_{\mbox{ext}}(\mathbf{r})-g_2 |\Psi|^2 + g_3|\Psi|^4 \right)\Psi,
	\end{equation}
	\noindent where $\Psi$ is the macroscopic wave function of the condensate, $V_{\mbox{ext}}(\mathbf{r})$ is the external trapping potential, $\hbar$ is the Planck constant, $M$ is the atomic mass, $g_2$ describes the two-particle interaction between atoms in the condensate and has the form 
 $	g_2=-4\pi \hbar^2 a_{s}/M$ where $a_s$ is the $s$-wave scattering length (positive for repulsive interatomic interaction and negative for attractive interaction). Parameter $g_3$ corresponds to the strength of three-particle repulsive interaction 
\cite{Dai2011,Ko2002,Braaten1999}. 
Here we take into account only conservative part of three-particle interaction. Thus Eq. (\ref{eq:zero}) conserves 
the norm of the condensate wave function $\Psi$ that is equivalent to the number of atoms:
\begin{equation}
    N = \int |\Psi|^2 d\mathbf{r}
\end{equation}
\noindent and energy:
	\begin{equation}
	\label{energy}
	E=\int\left(\dfrac{\hbar^2}{2M}|\mathbf{\nabla}\Psi|^2+V_{\mbox{ext}}|\Psi|^2-\dfrac{g_{2}}{2}|\Psi|^4+\dfrac{g_{3}}{3}|\Psi|^6 \right)d\mathbf{r}. 
	\end{equation}
	We consider a radially-symmetric time-independent harmonic trapping potential
	\begin{equation}
	\label{trap}
	V_{\mbox{ext}}(r,z)=\dfrac{M \omega^2_{\bot}}{2}r^{2}+\dfrac{M \omega_{z}^2}{2}z^{2},
	\end{equation}
	\noindent  where $r^{2}=x^{2}+y^{2},$  $\omega_{\bot}$ is the trap frequency in the transverse plane, $\omega_{z}$ is the longitudinal trapping frequency. For $\omega_{z}\gg\omega_{\bot}$ the condensate has a disk-shaped form (the longitudinal oscillator length is small comparing to the transverse size) and can be described by quasi two-dimensional (2D) wave function $\psi(r,t)$. We assume that the system is tightly confined in the $z$ direction 
	\begin{equation} 
	\label{separation}
	\Psi(r,z,t)= \psi(r,t)\Upsilon(z,t),
	\end{equation}
	where $\Upsilon(z,t)=(\pi l_{z})^{-1/2}\exp\left(-i\omega_{z}t/2 -z^2/2 l_{z}^2\right)$ is the ground state wave function in the oscillatory  potential $V_{z}(z) = M\omega_z^2z^2/2.$ Here $l_{z}=\sqrt{\hbar/(M\omega_{z})}$ is an oscillatory length in the $z$ direction. After integrating out the longitudinal coordinates in the Eq. (\ref{eq:zero}), we obtain 2D GPE considered in the following sections.
	\section{Dynamics of perturbed two-dimensional solitons in trapped BECs}\label{Sec_Trapped}
	In the model describing ultracold dilute atomic 2D BEC in radial harmonic trap $V_{\bot}(r)=M \omega^2_{\bot} r^{2}/2$ three particle interactions can be neglected $(g_3=0)$ and stationary soliton solutions exist both for attractive ($g_2>0$) and repulsive ($g_2<0$) two-particle interactions. 
	In terms of harmonic oscillator units [${t}\to t\omega_{\bot},$ ${r}\to r/l_{\bot},$ ${\Psi}\to{l_{\bot}\Psi },$ $g=\sqrt{8\pi}a_s/l_z,$ where  $l_{\bot}=\sqrt{\hbar/(M\omega_{\bot})}$] 2D GPE can be written in dimensionless form
	\begin{equation}
	\label{GPE}
	i \dfrac{\partial \psi}{\partial t}+\dfrac{1}{2}\left(\dfrac{d^2}{d r^2} + \dfrac{1}{r}\dfrac{d}{d r}\right) \psi- \dfrac{1}{2} r^2 \psi-g|\psi|^2\psi=0, 
	\end{equation}
	\noindent where $\psi$ is dimensionless wave-function and $g$ is the dimensionless 2D interaction constant, $r$  is the polar radius.
	Stationary soliton solution can be found as  $\psi(r,t)=e^{-i\mu t}\Psi(r)$ by solving equation
\begin{equation}
    \mu \Psi=\left [-\dfrac{1}{2}\left(\dfrac{d^2}{d r^2} + \dfrac{1}{r}\dfrac{d}{d r}\right)+\dfrac{1}{2} r^2 +g |\Psi|^2 \right ] \Psi,
\end{equation}
	 were $\mu$ is a chemical potential. Stationary solution characterized by constant density distribution so any voices cannot appear during evolution.
		But additional radially-symmetric perturbations in the initial condition can lead to the vortex formation.
	As perturbation we use
	\begin{equation}\label{eq:perturb}
	\Psi(r)\to\dfrac{1}{a} \Psi\left(\dfrac{r}{a}\right),
	\end{equation}
	where $a$ is a parameter of deformation. It's possible to stretch or extend the disc-shaped condensate in the radial direction by using external field (we only use radially-symmetric perturbations with $a>1$) and then release it at the moment of time $t=0$. 
	
\begin{figure}[H]	
		\begin{minipage}[]{0.52\linewidth}
			\center\includegraphics[width=1.1\linewidth]{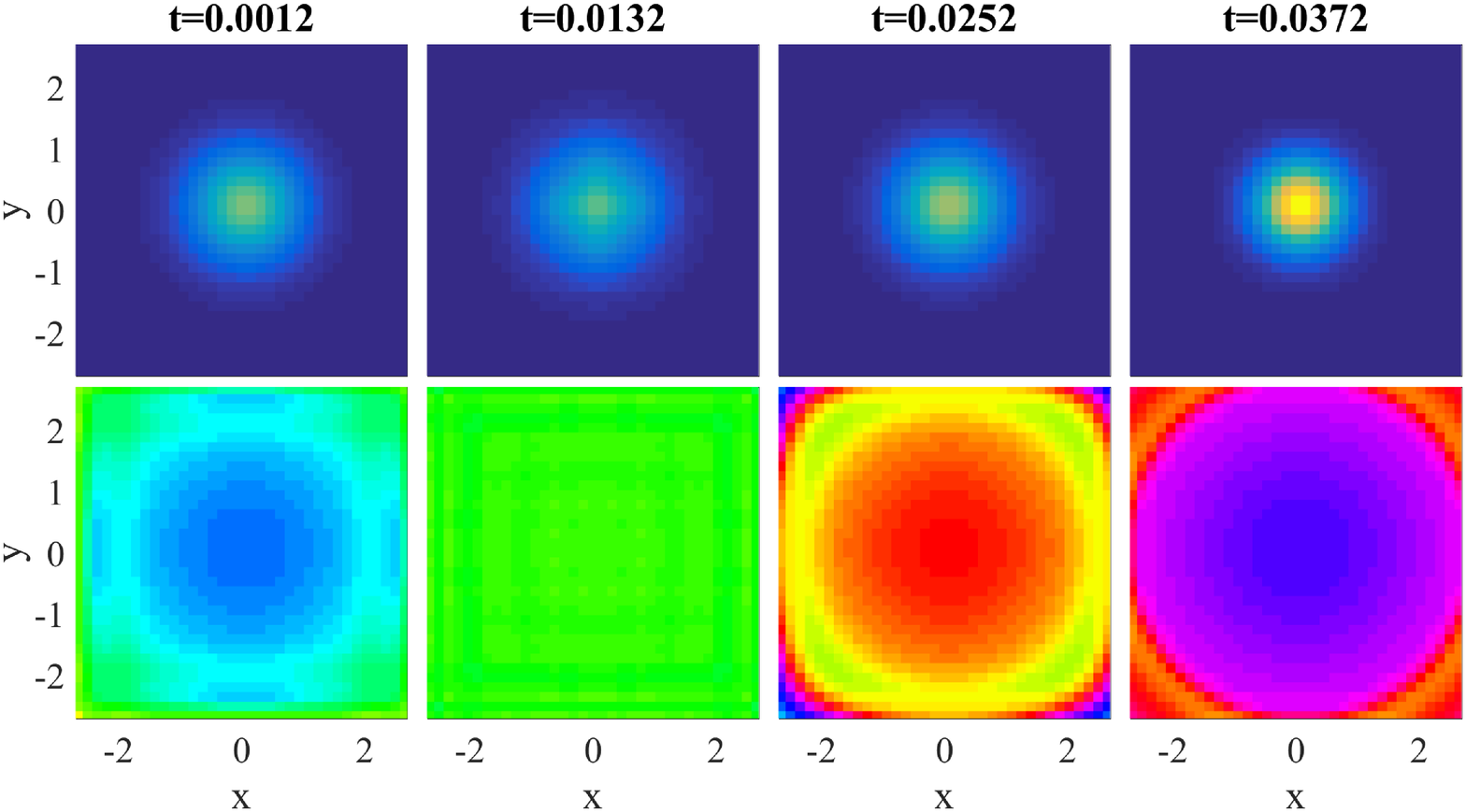}  (a) \\
		\end{minipage}
		\hfill
		\begin{minipage}[]{0.47\linewidth}
			\center{\includegraphics[width=1.1\linewidth]{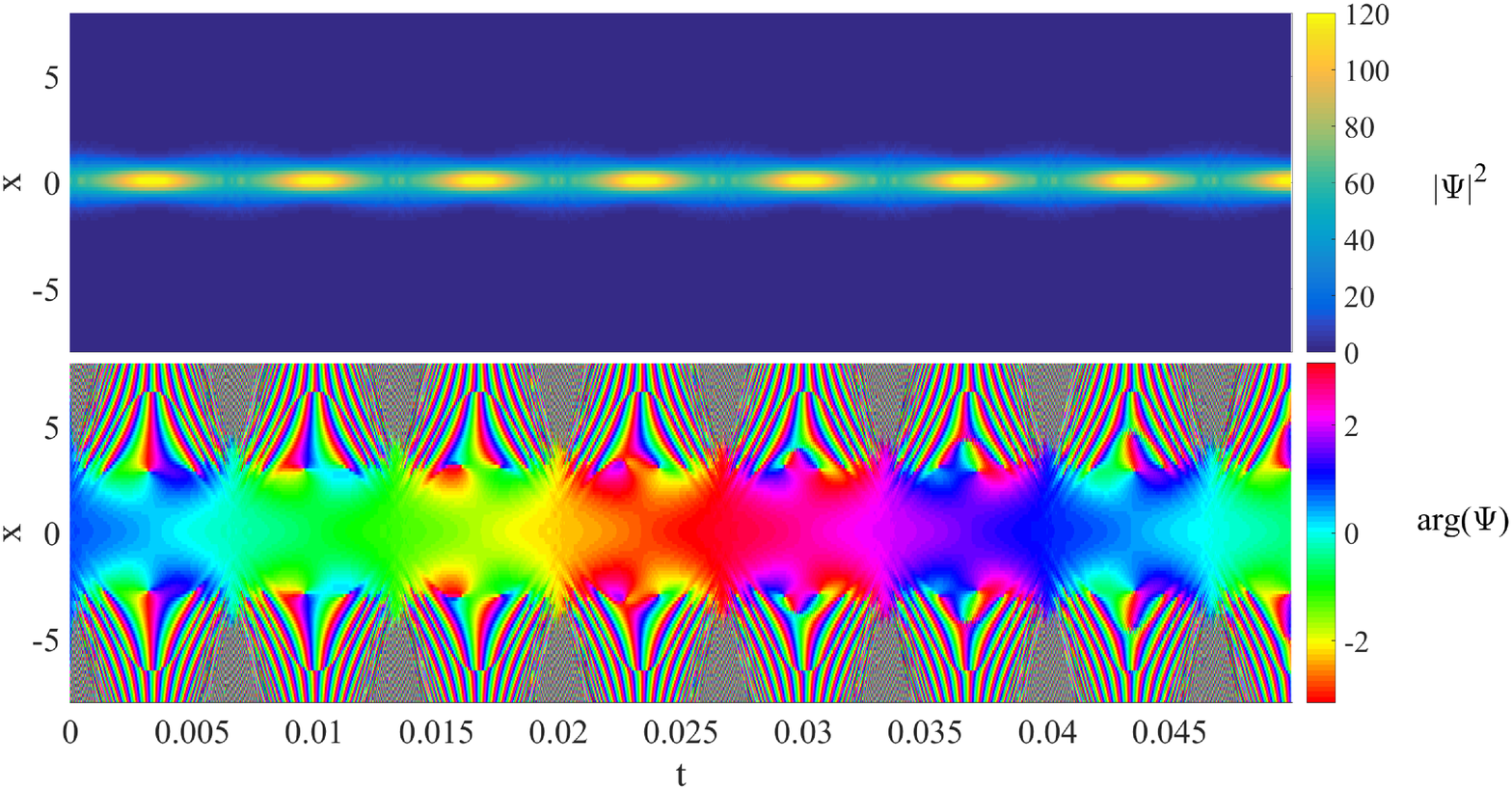} }\\(b)
		\end{minipage}
		\caption{  Dynamics of perturbed soliton in BEC with attractive two-particle  interaction for following parameters $g=-0.018,$ $N=175,$ $a=1.2$. (a) Dynamics of the density (top row) and phase (bottom row) of the condensate.   (b) Spatiotemporal distribution of the density (top row) and phase (bottom row) of the condensate. }	\label{fig:g1}
	\end{figure}

    We use the split-step (Fourier) method  \cite{Agarwal,XINRAN2017} to find numerically  solutions of Eq.  (\ref{GPE})  for different $N$. 	
	We consider BEC of $N=175$ of $^{23}$Na atoms in the  harmonic trap with the following parameters: $\omega_{\bot}=75 $ Hz,  $\omega_{z}=750$  Hz, $g=-0.018.$  
    An example of the time evolution of the condensate is shown in Fig. \ref{fig:g1} (a).
    Slice $y=0$ as a function of time is a spatiotemporal distribution of the condensate. An example of the spatiotemporal distribution of phase and density are shown in Fig.  \ref{fig:g1} (b).
	\begin{figure}[H]	
		\begin{minipage}[]{0.515\linewidth}
			\center\includegraphics[width=1.1\linewidth]{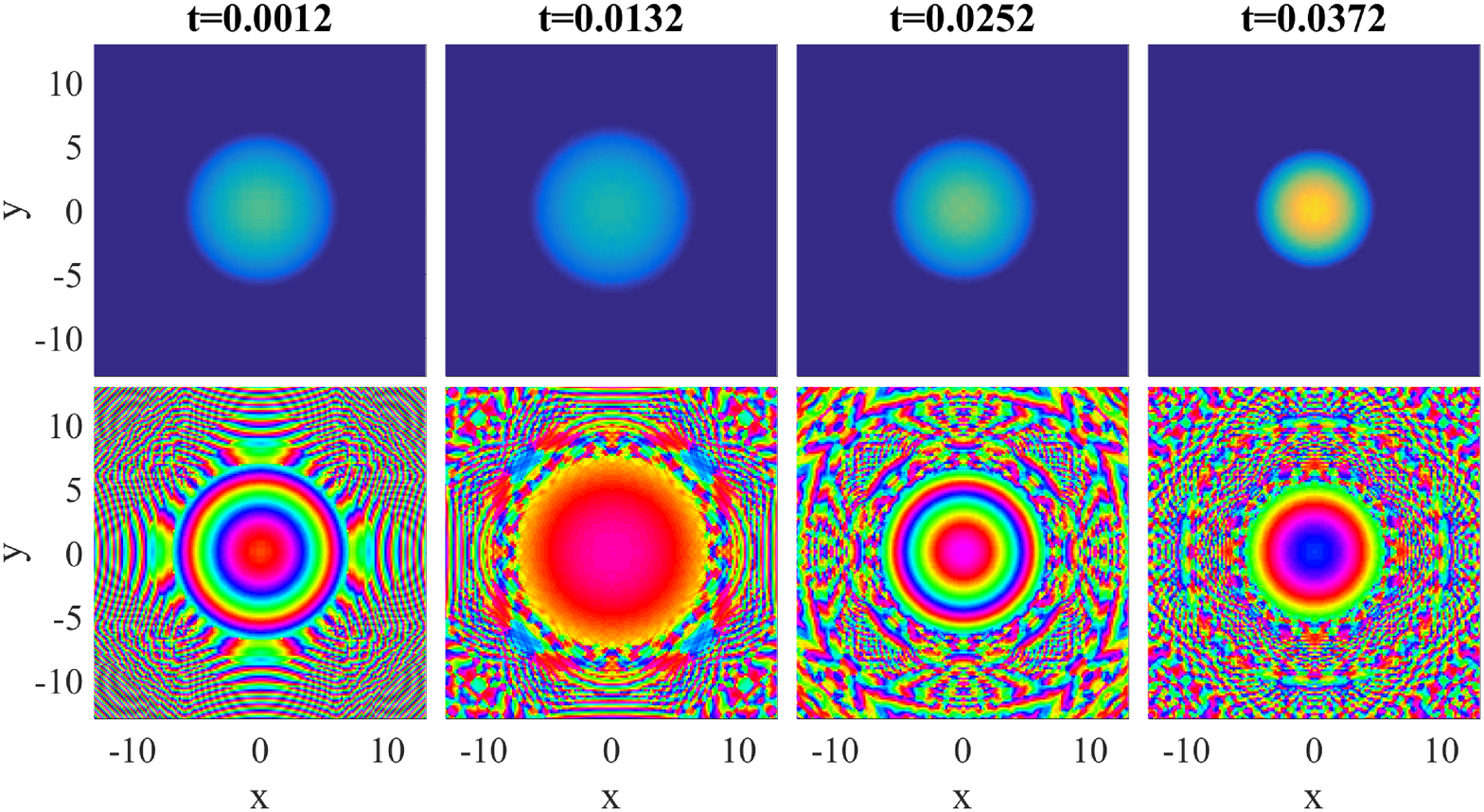}  (a) \\
		\end{minipage}
		\hfill
		\begin{minipage}[]{0.48\linewidth}
			\center{\includegraphics[width=1.1\linewidth]{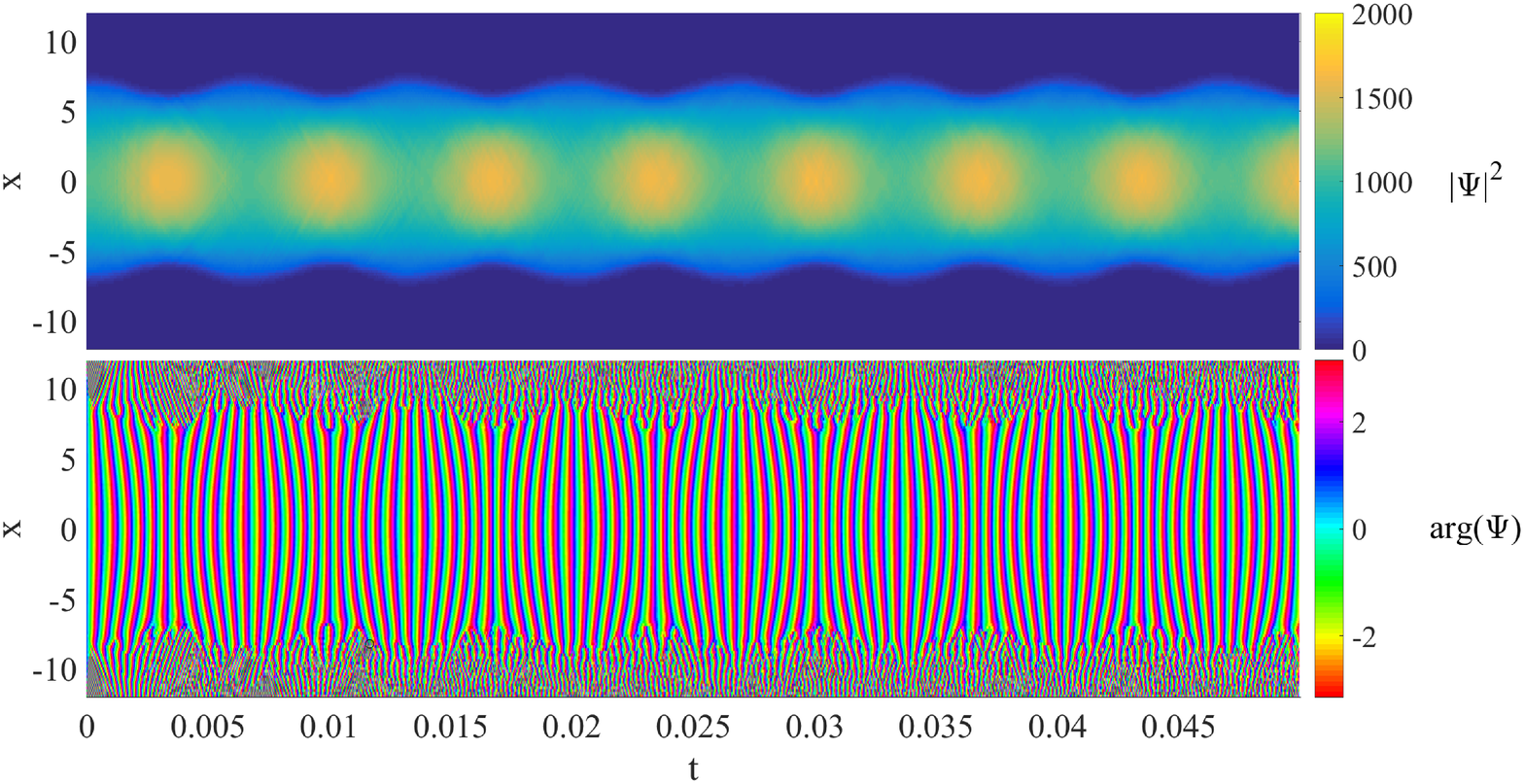} }\\(b)
		\end{minipage}
		\caption{ Dynamics of perturbed soliton in BEC with repulsive two-particle interaction for the following parameters $g=0.018,$ $N=10^{5},$ $a=1.1.$ (a) Dynamics of the density (top row) and phase (bottom row) of the condensate. (b) Spatiotemporal distribution of the density (top row) and phase (bottom row) of the condensate.}	\label{fig:g3}
	\end{figure}
	Vortices in a complex wavefunction are characterised by a phase circulation about the vortex core, and also by a positive curvature of the probability density. In small density regions wavefunction has little physical significance, but phase fluctuates wildly \cite{Jezek2009}. 
	We use the Wavelet denoising toolbox (Matlab) to clean the low-density region of numerical artifacts to consider phase structure. The vortex detector couldn't find any vortex core after the denoising procedure. It could therefore be stated that there were no spatiotemporal vortices in the system.  The same results for the systems with repulsive two-particle interaction (see Fig.  \ref{fig:g3} (a) and (b)). Various sets of system parameters has been checked but none of them didn't allow us to detect the spatiotemporal vortex.

	It is clear that external trapping potential plays a key role in the character of the system dynamics at the condensate low-density region. Thus, even strong perturbation ($a=2, 3, 5$) cannot lead to the vortex formation. 
	 However, an external potential is not necessarily  required for self-sustained condensates supported by nonlinear interactions in BEC. Various nonlinear effects can lead to formation of stable localized coherent structures as well. In the following section we consider atomic cloud without external trapping potential where attractive two-particle interactions are balanced by repulsive three-particle interactions.

	\section{Spatiotemporal vortex rings in a trapless BEC}\label{Sec_qubic_quintic}
The stationary solutions of the GPE Eq. (\ref{eq:zero}) without external potential are known to be unstable with respect to collapse for attractive  interaction and has no localized stationary solutions for repulsive interactions. Account on two-particle interactions and neglecting higher-order effects gives an adequate approximation for the dilute condensate trapped by external potential. However in the trapless condensate with attractive two-particle interactions higher-order effects, such as repulsive three-particle interactions, can play a crucial role since it can arrest a catastrophic collapse of the condensate. The model can be improved by adding a repulsive quintic nonlinear interaction term in Eq. (\ref{GPE}). The dimensionless 2D GPE then acquires the form
	
	
	\begin{equation}\label{eq:NLSE0}
	i\partial_{t}\Psi+D\left(\dfrac{d^2}{d r^2} + \dfrac{1}{r}\dfrac{d}{d r}\right)\Psi+g_2 |\Psi|^2 \Psi- g_3|\Psi|^4 \Psi=0,
	\end{equation}
	\noindent where 
	$D,$ $g_{2}$ and $g_{3}$ are dimensionless. The value of the three-body coupling constant $g_3$ should be small comparing to the two-particle coupling constant $g_2$. Such model  is well known as a model with cubic-quintic nonlinearity in nonlinear optics \cite{Davydova2004} and physics of ultracold atoms.

    We have the freedom of choice of $\omega_{z}$ so for simplicity we choose $\omega_{z}=1$ Hz.
	In the Eq. (\ref{eq:NLSE0}) with two-particle attractive and three-particle repulsive interaction both constants $g_2$ and $g_3$ are positive. Consider a change of coordinates ${\psi(r,t)} \to {\Psi(\rho, \tau)}$, where  $\Psi(\rho)=\psi(r)\sqrt{g_3/g_2},$
	$\rho=r\sqrt{g_2^2/(D g_3)},$ $\tau=t g_2^2/(\mu g_3),$ were 	$\mu$  is a dimensionless chemical potential.  Therefore Eq. (\ref{eq:NLSE0}) can be rewritten in form
	\begin{equation}\label{NLSE}
	i\partial_{\tau} \Psi +\left(\dfrac{d^2}{d \rho^2} + \dfrac{1}{\rho}\dfrac{d}{d \rho}\right)\Psi+|\Psi|^2\Psi-|\Psi|^4\Psi=0.
	\end{equation}
	
	It is also convenient to define a rescaled chemical potential $\lambda=\mu g_3/g_2^2$.  The real number of atoms (physical) can be found by  $N_{phys}=l_{z} N/(4\sqrt{2\pi}a_{s}),$  where $N$ is the norm of the new condensate wave function $\Psi(\rho,\tau)$.
	 To find a stationary state of Eq. (\ref{NLSE}), we write $\Psi(\rho,t)=\psi(\rho)\exp{(-i\lambda t)}:$
	 \begin{equation}\label{StatNLSE}
	     \lambda\Psi=\left[-\left(\dfrac{d^2}{d \rho^2} + \dfrac{1}{\rho}\dfrac{d}{d \rho}\right)-|\Psi|^2+|\Psi|^4    \right]\Psi.
	 \end{equation}
	We use the Thomas algorithm \cite{XINRAN2017} to find stationary solution of  Eq. (\ref{StatNLSE}). This algorithm allows us to find stable soliton solutions in the range of $\lambda$ from $-0.152$ to $0.$ The range can be extended using another method, for example, "shooting" method \cite{Davydova2004}. Dynamical simulations of the perturbed stationary state demonstrate the spatiotemporal vortex ring (STVR) formation for strong perturbations ($a>1.5$). And it is noteworthy that we have not observed  STVR in the systems with small perturbations.
	
\begin{figure}[H]
		\begin{minipage}[]{0.49\linewidth}
			\center\includegraphics[width=1\linewidth]{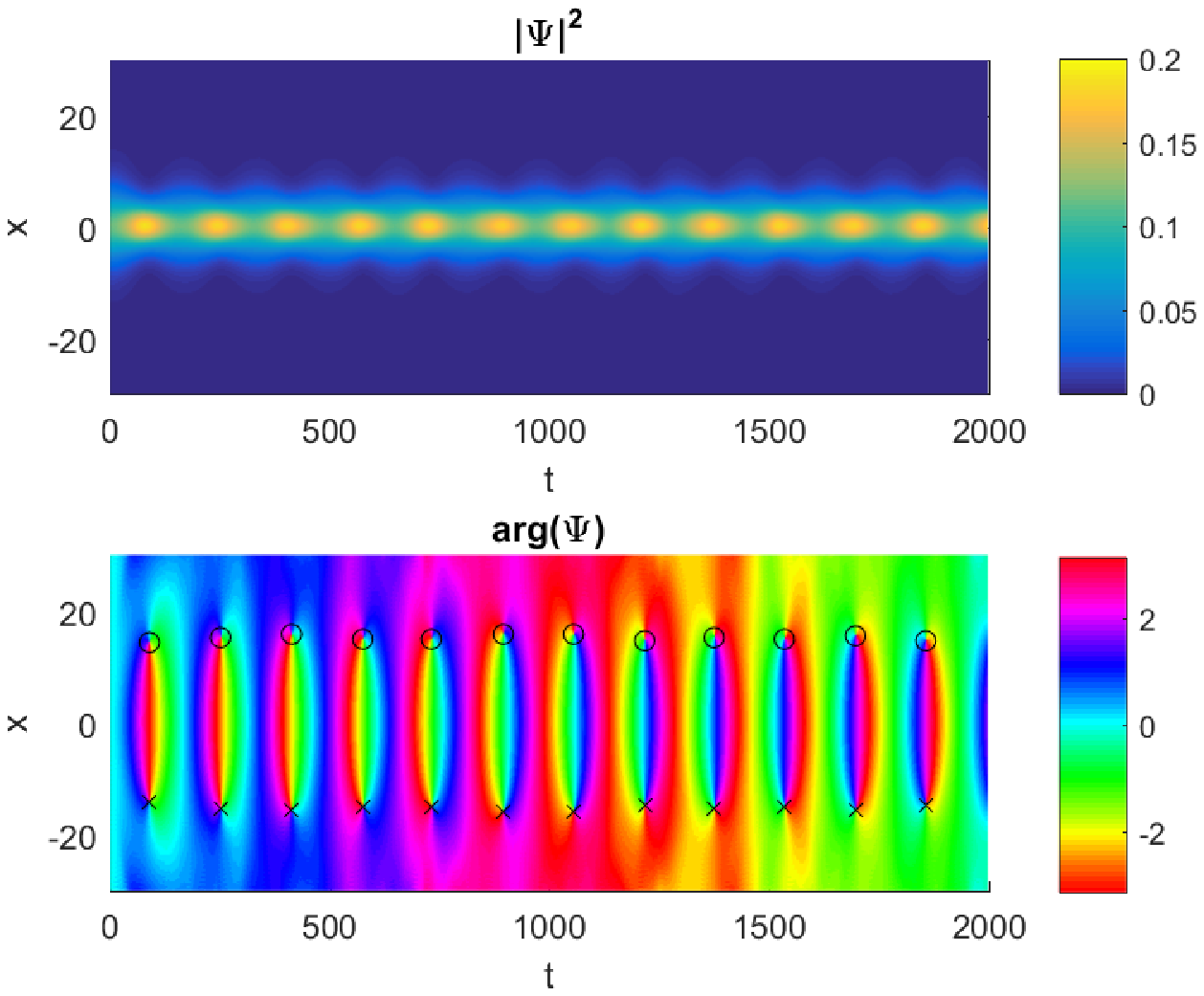}  	
			\\(a)
		\end{minipage}
		\hfill
		\begin{minipage}[]{0.49\linewidth} 
			\center\includegraphics[width=1\linewidth]{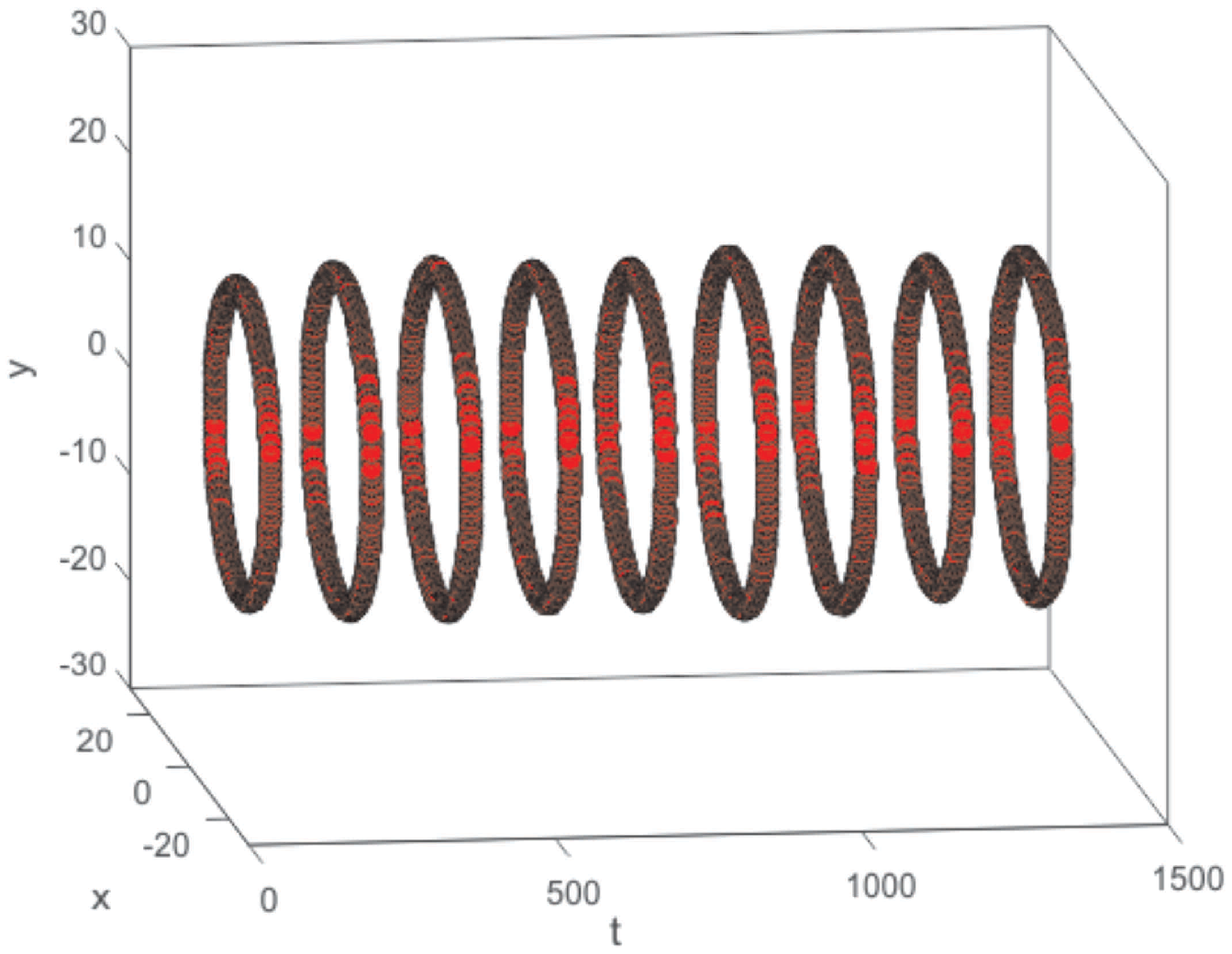} \\(b)
		\end{minipage}
		\caption{Dynamics of perturbed soliton in BEC with attractive two-particle interactions and three-particle repulsive interaction for the following parameters $\lambda=-0.05,$ $N=15,$ $a=1.5$ (a) Spatiotemporal distribution in slice $y = 0$ of the density (top row) and phase (bottom row) of the condensate (b) Spatiotemporal ring vortices.}
		\label{fig:lambda=-0.05}
    	\end{figure}

	The spatiotemporal vortex for small $|\lambda|$ is shown in Fig.  \ref{fig:lambda=-0.05}. By analogy with the previous section Fig.  \ref{fig:lambda=-0.05} (a) represents a slice $y = 0$ as a function of time. 
	At the equal time intervals two easily distinguishable points with phase winding appear. The one with a clock-wise winding and the other one with anticlock-wise winding which correspond to the vortex and antivortex responsibly. Combining this with data for other slices ($y \neq 0$) we build (2+1)D graph of all winding points in Fig. \ref{fig:lambda=-0.05} (b). It's remarkable that the radius of STVR is a periodic function of time when the Hamiltonian of this system isn't periodical. Therefore we can assume that continuous time-translation symmetry for this system is broken. STVR spontaneously appears in BECs so they can be related to the time crystals.

	Time crystals are time-periodic self-organized structures that were recently described in \cite{Shapere2012, Wilczek2012}. Those works raised a challenging issues of the existence of systems with spontaneously broken time-translation symmetry. In other words, how could a many-body system be self-organized in time to start a spontaneous periodic motion? It was anticipated that it is possible to prepare a many-body system in a state with an infinitesimally weak perturbation that will reveal periodic motion \cite{Wilczek2012,Sacha2018}. It was expected that time crystals can appear even in the lowest energy states, but it was shown recently that this idea cannot be realized \cite{Bruno2013,Watanabe2015}. However it was demonstrated experimentally that the discrete time translation symmetry can be spontaneously broken accompanied by discrete or Floquet time crystals nucleation \cite{Gibney2017, PhysRevLett.118.030401, Smits2018}.


\begin{figure}[H]
		\begin{minipage}[]{0.49\linewidth}
			\center\includegraphics[width=1\linewidth]{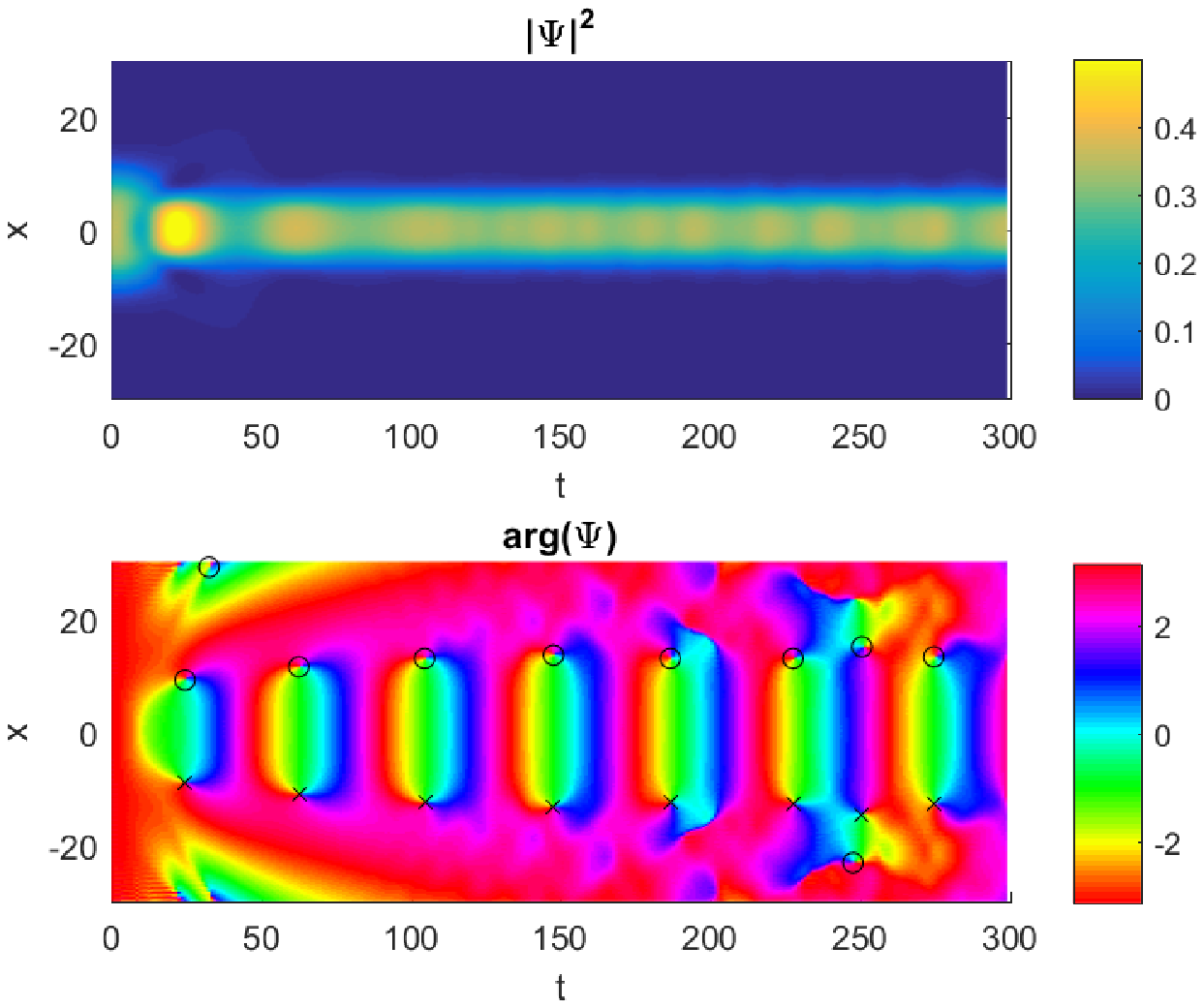}  	
			\\(a)
		\end{minipage}
		\hfill
		\begin{minipage}[]{0.49\linewidth} 
			\center\includegraphics[width=1\linewidth]{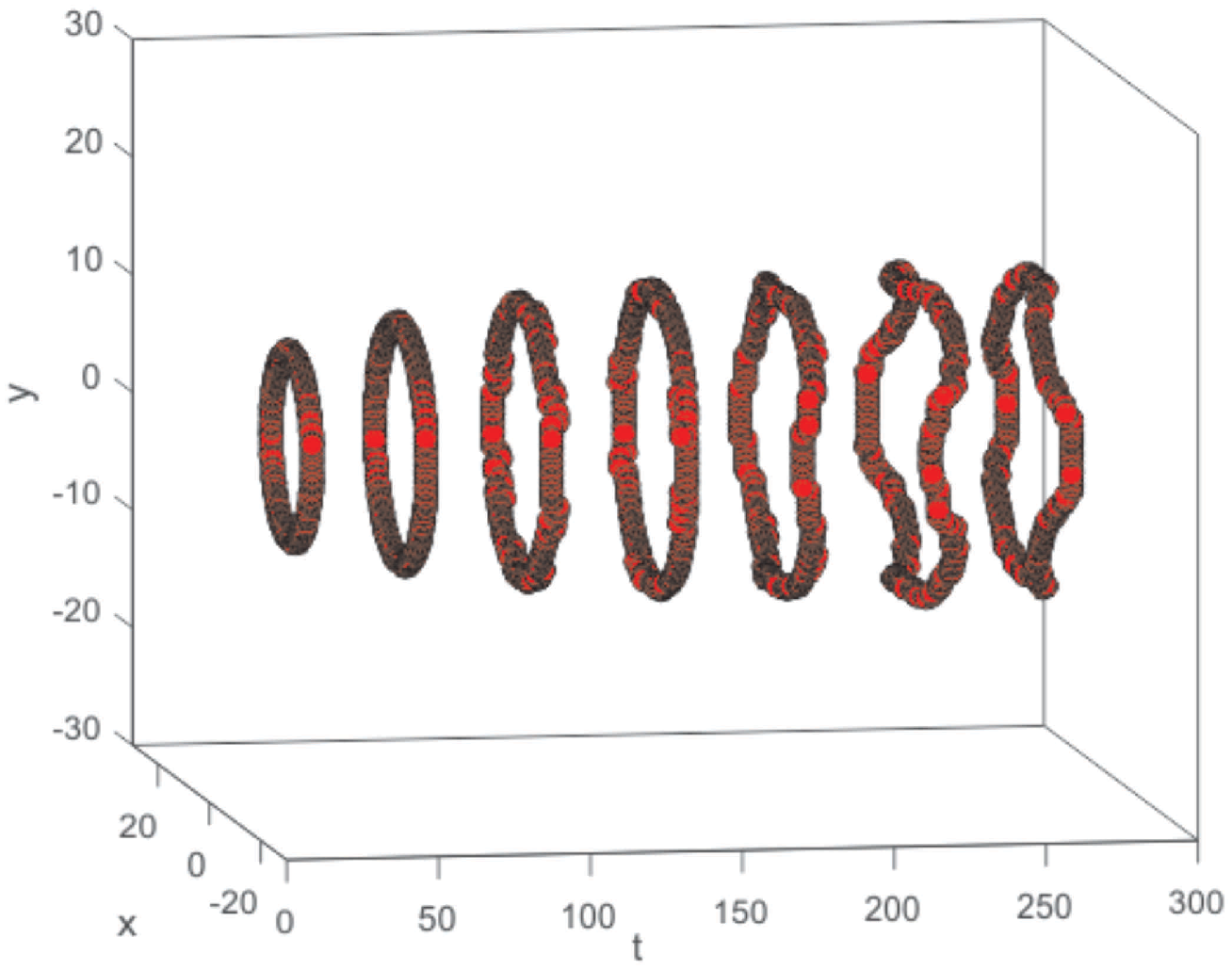} \\(b)
		\end{minipage}
		\caption{Dynamics of perturbed soliton in BEC with  attractive  two-particle  interactions  and  three-particle  repulsive  interaction  for the  following  parameters $\lambda =-0.152,$ $N=95$ and $a=1.5.$  (a) Spatiotemporal distribution in slice $y= 0$ of the density (top row) and phase (bottom row) ofthe condensate (b) Spatiotemporal ring vortices.}\label{fig:lambda=-0.125}
	\end{figure}

	The spatiotemporal vortices for bigger $|\lambda|$ (the bottom edge of the stable soliton solutions  range) are shown in Fig. \ref{fig:lambda=-0.125}. STVR have more complex form for these parameters and they are not localized in time anymore (see Fig.  \ref{fig:lambda=-0.125} (b)).

\begin{figure}[H]
		\begin{minipage}[]{0.48\linewidth}
			\center{\includegraphics[width=1\linewidth]{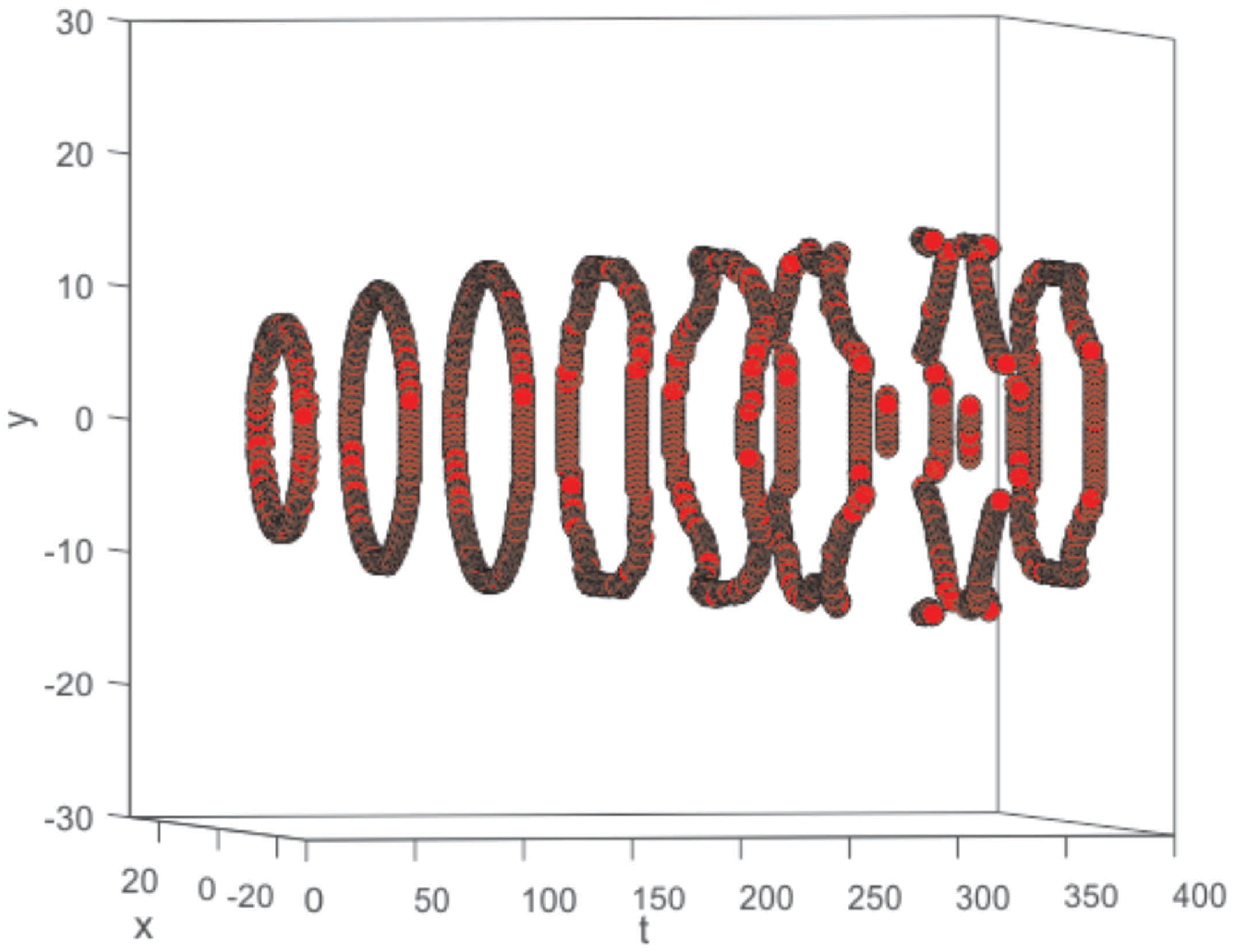} }  \\(a)
		\end{minipage}
		\hfill
		\begin{minipage}[]{0.48\linewidth}
			\center{\includegraphics[width=1\linewidth]{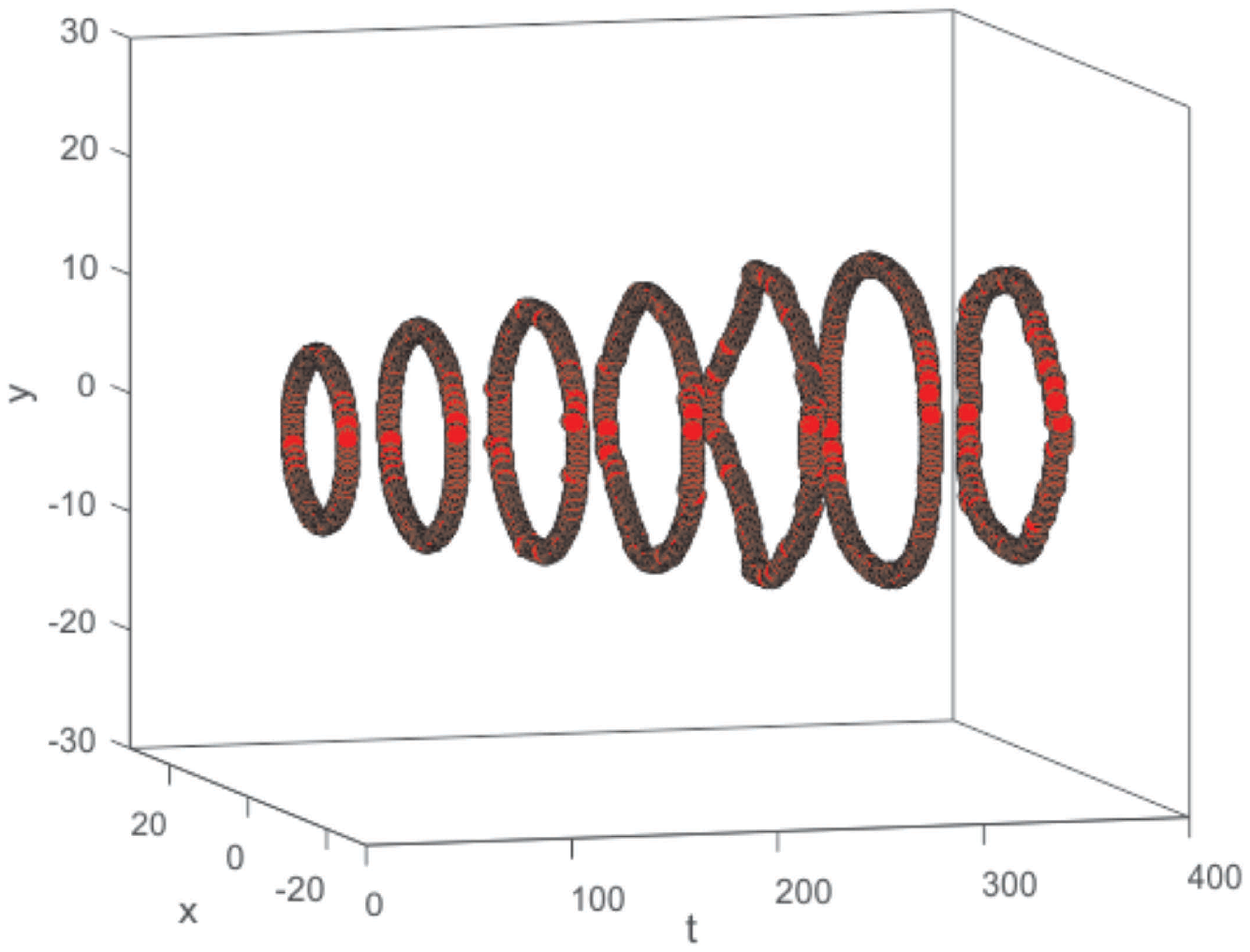} } \\(b)
		\end{minipage}
		\caption{  Spatiotemporal vortex rings for a systems with $\lambda =-0.125,$ $N=36.9$ and for (a)  $a=1.5$; (b)  $a=2.$}
		\label{rings_form}
	\end{figure}

	STVR form can be changed by increasing the perturbation parameter. For example, in  Fig.  \ref{rings_form} shown a stabilization of the vortex form to ring-like shape after increasing the perturbation parameter by one third. It is worth to emphasise that in the low-density region more complex systems can occur (small rings, vortex threads, etc) but it's very unlikely that they can be detected experimentally.

\begin{figure}[H]
		\centering
		\includegraphics[scale=0.7]{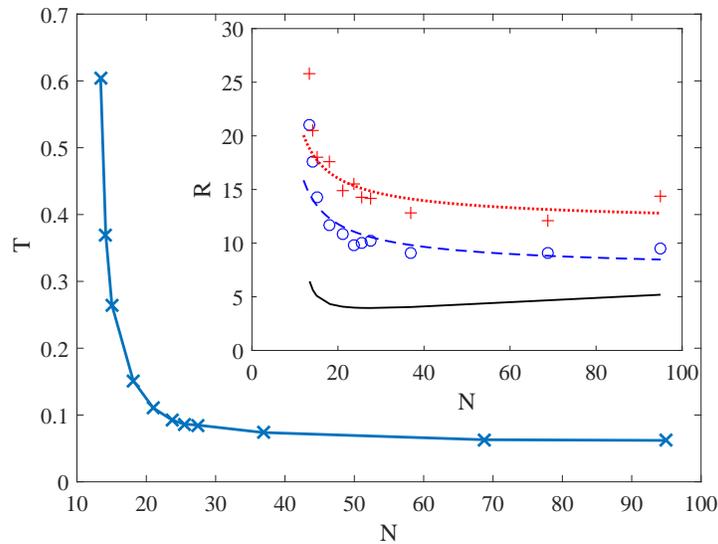}
		\caption{ The period of appearance of rings as a function of the number of particles for $ a = 1.5. $ Inset shows the radius of vortex rings as a function of the number of particles in the system at $ a =  1.5$. The black line corresponds to an effective radius of the soliton, blue dashed line with 'o' corresponds to the minimum radius of the rings in each simulation. The red dotted line with '+' corresponds to the maximum radius values.}
		\label{fig:g5}
	\end{figure}

	The ring's radius as a function of the numbers of particles in the condensate is shown in the inset in Fig.  \ref{fig:g5} for the same deformation  parameter $a =1.5$. The frequency of the STVR occurrences is increasing with increasing of the number of particles in the system (see Fig.  \ref{fig:g5}).
	
	One can see, that the least deformed rings is the one with the smallest radius (see Fig. \ref{rings_form}). The largest rings in the system are subject to significant fluctuations and, in some systems, have greatly deformed, up to the loss of their initial form and become transformed into other structures.
	\section{Conclusions}
	It was found the conditions for formation of spatiotemporal vortex rings in ultracold atomic gases. These vortex structures exhibit  phase dislocation both in space and time. A sequence of spatiotemporal vortex rings appears as a periodical in time edge phase dislocation at the low-density region of a perturbed atomic Bose-Einstein condensate. 	
	
	The dynamics of perturbed stationary soliton solutions of the Gross-Pitaevskii equation for two-dimensional condensate in the external trapping potential is investigated using numerical simulations. Both attractive and repulsive interparticle interactions are studied. It turns out that no spatiotemporal vortex rings can be detected when the trapping potential strongly suppress the condensate density at the periphery of the atomic cloud.
	
	It is revealed that the  spatiotemporal vortex rings can be trustworthy observed in the system without the external trapping potential supported by competing attractive two-particle and repulsive three-particle interactions. The sequence of spatiotemporal vortex rings for systems with different initial perturbations and number of particles are studied. It is found that the temporal period of the sequence of spatiotemporal vortex rings is mostly determined by number of particles, while the ring radius depends both on amplitude of the deformation and on the number of particles. 
	
	We hope that developed in this work theoretical findings open up the perspective for experimental observation of novel type of topological coherent structures in ultracold gases.
	
	\section{Acknowledgements}	

O. C. thanks to M.  Shchedrolosiev for useful discussion.

\end{document}